\documentclass[pdflatex,sn-mathphys,Numbered]{sn-jnl}

\usepackage{siunitx} 
\usepackage[inline]{enumitem} 
\usepackage{subcaption} 
\usepackage{upgreek} 

\usepackage{graphicx}%
\usepackage{multirow}%
\usepackage{amsmath,amssymb,amsfonts}%
\usepackage{amsthm}%
\usepackage{mathrsfs}%
\usepackage[title]{appendix}%
\usepackage{xcolor}%
\usepackage{textcomp}%
\usepackage{manyfoot}%
\usepackage{booktabs}%
\usepackage{algorithm}%
\usepackage{algorithmicx}%
\usepackage{algpseudocode}%
\usepackage{listings}%

\newcommand{\lRes}{
    l_\mathrm{res}
}
\newcommand{\epsEff}{
    \varepsilon_\mathrm{eff}
}
\newcommand{\Qi}{
    Q_\mathrm{i}
}
\newcommand{\Ql}{
    Q_\mathrm{l}
}
\newcommand{\Qc}{
    Q_\mathrm{c}
}
\newcommand{\df}{
    \mathrm{\partial} f
}

\DeclareSIUnit\sq{\Box}

\begin{document}

\title[Directional Filter Design and Simulation for Superconducting On-chip Filter-banks]{Directional Filter Design and Simulation for Superconducting On-chip Filter-banks}

\author*[1,2]{\fnm{Louis H.} \sur{Marting}}\email{l.h.marting@tudelft.nl}

\author[2]{\fnm{Kenichi} \sur{Karatsu}}\email{k.karatsu@tudelft.nl}

\author[1]{\fnm{Akira} \sur{Endo}}\email{a.endo@tudelft.nl}

\author[1,2]{\fnm{Jochem J. A.} \sur{Baselmans}}\email{j.baselmans@sron.nl}

\author[1,2,3]{\fnm{Alejandro} \sur{Pascual Laguna}}\email{alejandro.pascual@cab.inta-csic.es}

\affil*[1]{\orgdiv{Faculty of Electrical Engineering, Mathematics and Computer Science}, \orgname{Delft University of Technology}, \orgaddress{\street{Mekelweg 4}, \city{Delft}, \postcode{2628 CD}, \state{Zuid-Holland}, \country{The Netherlands}}}

\affil[2]{\orgname{SRON--Netherlands Institute for Space Research}, \orgaddress{\street{Niels Bohrweg 4}, \city{Leiden}, \postcode{2333 CA}, \state{Zuid-Holland}, \country{The Netherlands}}}

\affil[3]{\orgdiv{Centro de Astrobiología}, \orgname{CSIC-INTA}, \orgaddress{\street{Torrejón de Ardoz}, \city{Madrid}, \postcode{28850}, \country{Spain}}}

\abstract{Many superconducting on-chip filter-banks suffer from poor coupling to the detectors behind each filter. 
This is a problem intrinsic to the commonly used half wavelength filter, which has a maximum theoretical coupling of 50\,\%. 
In this paper we introduce a phase coherent filter, called a directional filter, which has a theoretical coupling of 100\,\%.

In order to to study and compare different types of filter-banks, we first analyze the measured filter frequency scatter, losses, and spectral resolution of a DESHIMA 2.0 filter-bank chip. 
Based on measured fabrication tolerances and losses, we adapt the input parameters for our circuit simulations, quantitatively reproducing the measurements. We find that the frequency scatter is caused by nanometer-scale line-width variations and that variances in the spectral resolution is caused by losses in the dielectric only.

Finally, we include these realistic parameters in a full filter-bank model and simulate a wide range of spectral resolutions and oversampling values. 
For all cases the directional filter-bank has significantly higher coupling to the detectors than the half-wave resonator filter-bank. 
The directional filter eliminates the need to use oversampling as a method to improve the total efficiency, instead capturing nearly all the power remaining after dielectric losses.}


\maketitle

\section{Introduction}
On-chip superconducting filter-bank spectrometers are a promising technology for wideband spectroscopic observations in the (sub-)millimeter wavelength range \citep{endoFirstLightDemonstration2019,redfordSuperSpecOnChipSpectrometer2022,cataldoSecondgenerationMicroSpecCompact2019}. 
However, the efficiency of current on-chip filter-banks is poor. 
Partly this is due to dielectric loss in the filters, but more fundamentally the current designs based on shunted half-wave resonator filters have a maximum theoretical coupling to the detector behind each filter of \qty{50}{\%} \cite{pascuallagunaTerahertzBandPassFilters2021,kovacsSuperSpecDesignConcept2012,endoDesignIntegratedFilterbank2012}. 
Here, we introduce a phase-coherent filter, called a directional filter, as an alternative filter design that ideally has an efficiency of \qty{100}{\%}.

We model the performance of a directional filter numerically and compare it to the standard shunted half-wave resonator. 
This is done for an individual filter and for a full filter-bank.  
Additionally, we study the effects of fabrication tolerances and dielectric losses for a filter-bank with a spectral resolution of $R = f_0/\Delta f_0 = 500$ in a frequency band of \qtyrange{220}{440}{GHz}, which is the target design for the DESHIMA 2.0 instrument \citep{taniguchiDESHIMADevelopmentIntegrated2022}, soon to be fielded to the ASTE telescope in Chile. 
In our model we use the measured losses as well as the measured variations in losses and resonance frequency of the DESHIMA 2.0 filter-bank as input. Finally, a comparison is made between filter-banks using either directional filters or half-wave resonator filters for a variety of spectral resolutions and oversampling values.

In Section~\ref{sec:directional filter} we introduce the directional filter and show its performance. In Section~\ref{sec:filter-bank model} we show the results of simulations of  a filter-bank model with ideal components.
In Section~\ref{sec:tolerance analysis} we introduce fabrication tolerances and dielectric loss.
In Section~\ref{sec:filter-bank simulations} we study these tolerances and losses in realistic filter-banks across a range of spectral resolutions and oversampling values with full filter-bank simulations.

\section{Directional Filter}\label{sec:directional filter}
A directional filter \citep{cohnDirectionalChannelSeparationFilters1956,coaleTravelingWaveDirectionalFilter1956,oliverDirectionalElectromagneticCouplers1954,jonesCoupledStripTransmissionLineFiltersDirectional1956,tuttlePracticalDesignStripTransmissionLine1959} is a symmetric four-port device that acts as a channel separating filter. 
Directional filters are characterized by four basic properties \citep{cohnDirectionalChannelSeparationFilters1956}:
\begin{enumerate*}[label=\arabic*)]
    \item they have four ports, one of which is always isolated from the input port;
    \item the input power is filtered to one port with a band-pass response, while the remaining power emerges from another port with a complementary band-stop response;
    \item the input port is non-reflecting when the other ports are impedance-matched;
    \item these properties apply no matter which of the four ports is used.
\end{enumerate*}

\begin{figure}[htbp]
    \centering
    \subcaptionbox*{}{%
        \includegraphics[width=0.52\textwidth]{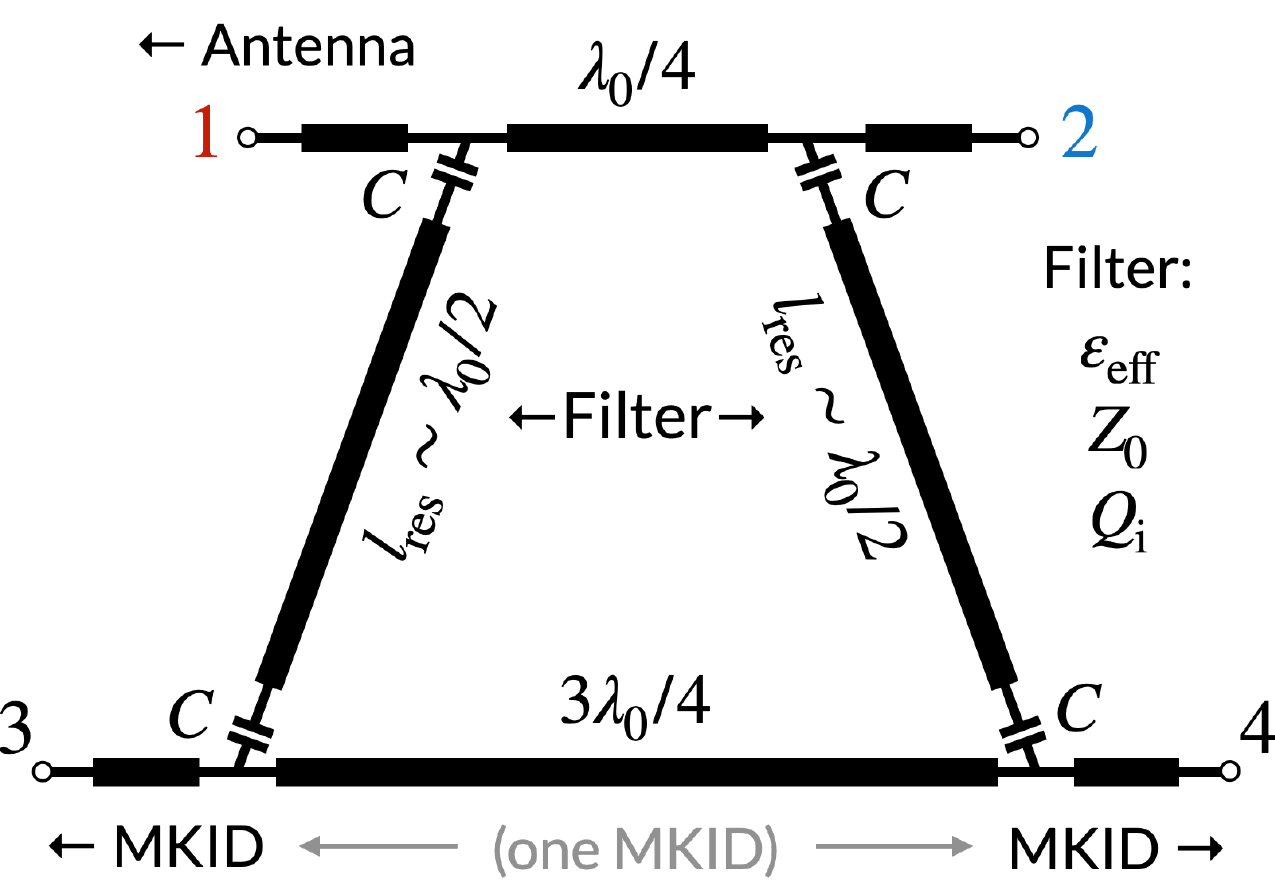}%
        \vspace{10 mm} 
    }%
    \subcaptionbox*{}{%
        \includegraphics[width=0.50\textwidth]{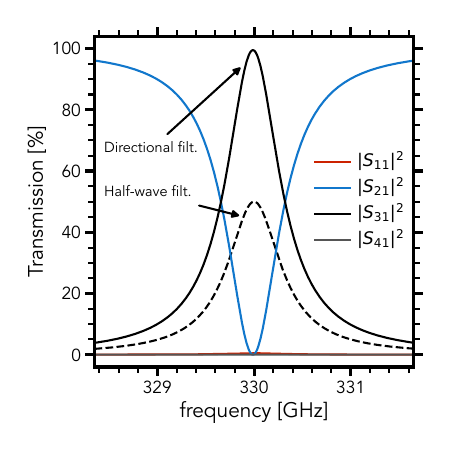}%
    }%
    \captionsetup{skip=-5mm}
    \caption{A lossless directional filter with resonance frequency $f_0$ at \qty{330}{\GHz}. 
    The circuit schematic is shown on the left. 
    The bottom transmission line is part of the MKID detector, for details we refer to the text.
    On the right, the filter S-parameters around the resonance frequency are shown. The numerical response is calculated by building a circuit model similar to Pascual Laguna et al. \cite{pascuallagunaTerahertzBandPassFilters2021}.}
    \label{fig:filter schematic}
\end{figure}

A schematic of a directional filter, which consists of two half-wave resonators capacitively coupled to transmission lines at the top and bottom, is shown in the left panel of Figure~\ref{fig:filter schematic}. 
The half-wave resonators are separated by a quarter wavelength at the top and by three-quarter wavelengths at the bottom.
By making port 1 be the input: port 2 gives a band-stop response; port 3 has a band-pass response; and port 4 is isolated. 
In this configuration, the top transmission line, which we call the signal line, carries the wideband input signal captured by a broadband antenna. 
This configuration is as easily fabricated as two half-wave resonators next to each other. The only additional requirements are that there is enough room for each coupler and that the bottom transmission line has a meander to create the additional half-wavelength. Both conditions are easily satisfied.

The bottom transmission line is part of the microwave kinetic inductance detector (MKID), where the sub-millimeter radiation is absorbed and detected. 
The quarter wavelength and three-quarter wavelength sections could be swapped without changing the function of the directional filter due to the symmetry of the structure. 
This configuration is chosen to reduce the signal line length, and thus its loss. 
Daisy-chaining ports 1 and 2 of separately tuned filters creates a filter-bank, as we shall do in Section~\ref{sec:filter-bank model}.

The right panel of Figure~\ref{fig:filter schematic} shows the numeric response of the directional filter around its resonance, for comparison the $|S_{31}|^2$ for a half-wave resonator filter is also given.
Shunted half-wave resonator filters cause an impedance mismatch on-resonance, which results in a theoretical transmission limit of \qty{50}{\%} to the detector behind each filter. 
The directional filter overcomes this limit by matching the incoming and outgoing signal using a phase-coherent structure. 
The maximum on-resonance transmission of an ideal directional filter is instead \qty{100}{\%}.

To absorb the power in the aluminium (Al) section of a $\lambda/4$ niobium titanium nitride (NbTiN)/Al hybrid MKID \citep{janssenHighOpticalEfficiency2013}, we integrate the hybrid section of the MKID with the bottom of the coherent filter circuit by coupling co-planar waveguide (CPW) lines with an Al central line to ports 3 and 4. 
These must have sufficient length to absorb all power in the Al central line. 
In this case both ports 3 and 4 see the characteristic impedance of the line, allowing to match both ports. 
This satisfies the third basic property of a directional filter.
The hybrid CPW line to be connected to port 4 is in series with a wide NbTiN CPW section which is parallel-coupled to the microwave readout line. The equally long hybrid CPW MKID section to be connected to port 3 is short-circuited to ground to set the other boundary condition establishing the $\lambda/4$ microwave resonance of the MKID.

\section{Filter-bank model}\label{sec:filter-bank model}
We will now discuss the performance of a filter-bank consisting of directional filters using a numerical model based on the model developed by Pascual Laguna et al. \cite{pascuallagunaTerahertzBandPassFilters2021}. 
For the filter-bank model we use ideal components and a filter overlap at \qty{50}{\%} of the peak response. 
The target design is a spectral resolution of $R = f_0/\Delta f_0 = 500$ in a \qtyrange{220}{440}{\GHz} band, which is the target design for the DESHIMA 2.0 instrument \citep{taniguchiDESHIMADevelopmentIntegrated2022}. 
This configuration is a choice and other configurations are also possible \citep{hailey-dunsheathStatusSuperSpecBroadband2014}, but we present our results in the most general way possible.

\begin{figure}[H]
    \includegraphics[width=\textwidth]{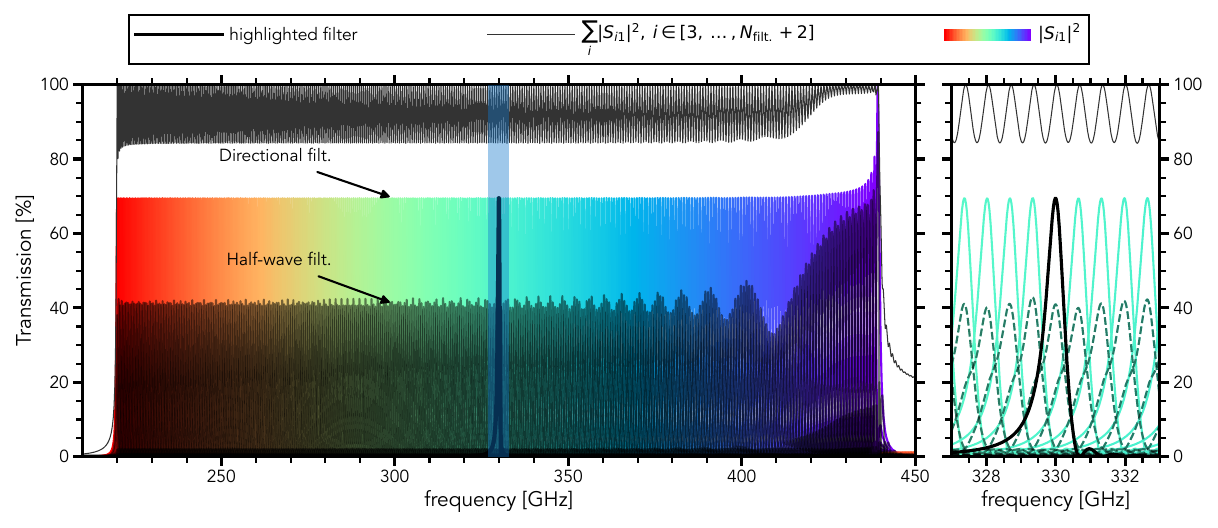}
    \caption{Response of a lossless filter-bank with directional filters (solid lines) and half-wave resonator filters (dashed lines). Color lines indicate the response of each channel. The thin black line indicates the total power absorbed by the directional filter-bank. The thick black line highlights an individual filter.}
    \label{fig:filter-bank}
\end{figure}

Figure~\ref{fig:filter-bank} shows the spectral response of a lossless filter-bank consisting of directional filters. 
Across the filter-bank operating band, the total aggregate power captured by all channels varies from \qtyrange{85}{100}{\%}, being maximum at the resonance frequency of each individual filter.
The average peak response of individual filters is around \qty{70}{\%}. Comparatively, the peak response of half-wave resonator filters is only \qty{40}{\%}.
Hence, the directional filter is almost twice as efficient in a filter-bank configuration compared to a half-wave resonator filter under ideal conditions (lossless and nominal design parameters) and no oversampling. 
Oversampling is defined as the ratio of the full-width half-maximum channel width over the channel resonance frequency separation, $\Delta f_i / (f_i - f_{i-1})$. 

The highest frequency filter, which is first in the signal line from the antenna, shows the \qty{100}{\%} efficiency that we expect from the single filter model.
The other filters settle at an efficiency of \qty{70}{\%}.
This is a direct result of the \qty{50}{\%} overlap design choice (oversampling = 1), where the tails of neighboring filters absorb the remaining \qty{30}{\%}.
This is further evidenced by the asymmetry of the filter shape in the zoom-in in Figure~\ref{fig:filter-bank} and by the \qty{100}{\%} of the \emph{total} power being absorbed ($\sum_i{|S_{i1}|^2},\: i \in [3, \dots, N_\mathrm{filt.}+2]$) at each filter's resonance frequency. 
It shows that power at those frequencies is neither reflected at the filter-bank input port nor dissipated at the load terminating the signal line.
Even at the intersection between adjacent filters the total power coupled is \qty{85}{\%}, significantly more than the value of \qty{60}{\%}  reported for  the half-wave resonator filter (see Pascual Laguna \cite{pascuallagunaOnChipSolutionsFuture2022}, Fig. 3.7).

\section{Tolerance analysis}\label{sec:tolerance analysis}
\begin{figure}[htb]
    \centering
    \includegraphics[width=0.8\textwidth]{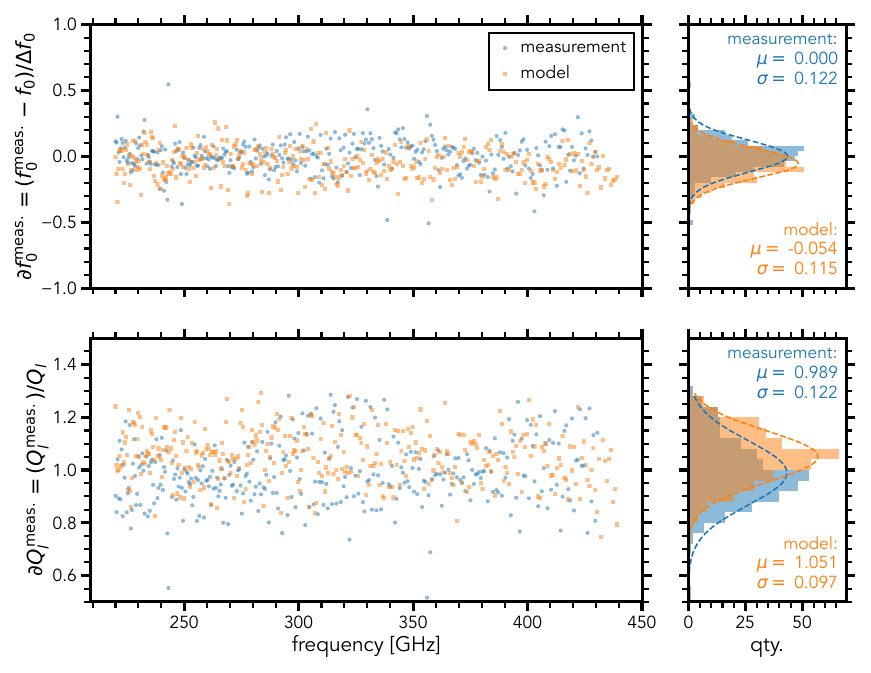}
    \caption{Normalized scatter of $\partial \Ql^\mathrm{meas.}$ and $\partial f_0^\mathrm{meas.}$ for a DESHIMA 2.0 chip with design values $\Ql = 500$ and $f_0 = \qtyrange{220}{440}{\GHz}$. $\Delta f_0 = f_0 / \Ql$. The blue points indicate the measured data, the orange points indicate our model results, we refer to the text for details. The offsets in the measurements have been normalized to only show scatter.}
    \label{fig:scatter_data}
\end{figure}

Using data from a half-wave resonator filter-bank for DESHIMA 2.0, we can include representative fabrication tolerances and dielectric losses in the filter-bank model.
The blue data in Figure~\ref{fig:scatter_data} shows these variances measured in the resonance frequency $f_0$ and loaded quality factor $\Ql$ for the DESHIMA 2.0 chip. 
The loaded quality factor $\Ql$ is equal to the spectral resolution $R = f_0 / \Delta f_0$, i.e. an oversampling value of 1.
The normalized $f_0$ scatter is defined as: $\partial f_0^\mathrm{meas.} = (f_0^\mathrm{meas.} - f_0) / \Delta f_0$.
The normalized $\Ql$ scatter is defined as: $\partial \Ql^\mathrm{meas.} = (\Ql^\mathrm{meas.})/\Ql$.
The first step in our analysis is to find which parameters in our model are responsible for the observed scatter in $f_0$ and $\Ql$.

Each resonator in the directional filter model in Figure~\ref{fig:filter schematic} has five input parameters, namely $\lRes$, $\epsEff$, $C$, $Z_0$ and $\Qi$.
Two parameters vary for each resonator in our filter-bank design: $\lRes$ is the resonator length; and $C$ is the coupling capacitance.
They are used to set the resonance frequency $f_0$ and coupling quality factor $\Qc$.
The other parameters, $\epsEff$, $Z_0$ and $\Qi$, define the resonator transmission line and are given by its cross-sectional dimensions and the material properties. 
They ideally stay constant across the filter-bank.

By taking a single resonator and varying each of these parameters slightly we can find how strongly each of them affects $f_0$ and $\Ql$.
In turn, we can model with SONNET how strongly each of these parameters is affected by realistic fabrication tolerances.
The result, reported in Table~\ref{tab:tolerance analysis}, is a set of model input parameters with their sensitivities mapped and linked to various underlying fabrication errors.
The table gives the deviation for each parameter required to create a change in $\Ql$ or $f_0$ of \qty{12}{\%} for a single resonator, which is the measured variance in DESHIMA 2.0 (see Figure~\ref{fig:scatter_data}).

It should be noted that the parameters are affected by multiple causes (e.g. variations in gap width of the couplers, line width, thickness of the materials, etc.), which were all investigated. 
However, only the most sensitive cause affecting the parameters is reported.

\begin{table*}[htb]
    \centering
    \caption{Mapping of our model parameters to fabrication errors. The last two columns give the error needed to achieve the offset in the column heading. The offset corresponds to the normalized variance of the measured DESHIMA 2.0 data in Figure~\ref{fig:scatter_data}. The bold parameters are used to add variances to the model. An empty value means that the parameter has no influence on that variance.}
    \label{tab:tolerance analysis}
    \begin{tabular}{l l c c}
        \toprule
        {Parameter}& {Cause} & {$\df_0^\mathrm{meas.} = \qty{12}{\%}$}   &   {$\mathrm{\partial}\Ql^\mathrm{meas.} = \qty{12}{\%}$}  \\
        \midrule
        $\lRes$    & length        & \qty{42}{\nm}   & \qty{17.7}{\um} \\
        $\epsEff$  & $R_\mathrm{s} \rightarrow  L_\mathrm{k}$      & \qty[per-mode=symbol]{0.2}{\pico\ohm\per\sq}  & \qty[per-mode=symbol]{113}{\pico\ohm\per\sq} \\
        $C$        & gap width        & \qty{5.5}{\nm}  & \qtyrange{20}{60}{\nm} \\
        {\boldmath $Z_0$}      & \textbf{line width}       & {\boldmath \qty{1}{\nm}\footnotemark[1]}  &  \\
        {\boldmath $\Qi$}      & \textbf{dielectric loss}  &  & {\boldmath \qty{23.4}{\%}} \\
        \bottomrule
        \multicolumn{4}{l}{ {\footnotesize \footnotemark[1] \qty{1}{\nm} corresponds to a variance of $\partial Z_0/Z_0 = \qty{1.2}{\%}$ in our model} }
    \end{tabular}

\end{table*}

From Table~\ref{tab:tolerance analysis} we can conclude that the scatter in $\Ql$ can \emph{only} be explained by $\Qi$ scatter as any other parameter causes a scatter in $f_0$ well exceeding the measured value in DESHIMA 2.0.
To understand why $\Qi$ scatter affects $\Ql$ so much, consider the following equation:
\begin{equation} \label{eq:Ql}
    \frac{1}{\Ql} = \frac{2}{\Qc} + \frac{1}{\Qi}.
\end{equation}
This equation gives $\Ql$ in terms of $\Qc$ and $\Qi$, assuming two equal couplers on each end of the resonator (hence the factor 2). 
From \ref{eq:Ql} it is clear that the sensitivity of $\Ql$ to $\Qi$ largely depends on the ratio between $\Qc$ and $\Qi$. 
If $\Qi$ is relatively small compared to $\Qc$, any variance on it will have a large impact on $\Ql$.
The average measured $\Qi$ of this chip is approximately $\num{1200}$, which is close to the $\Qc$ values of the measured chip.

The frequency scatter is most likely caused by a variance in $Z_0$, caused by nanometer scale line width variations from filter to filter. 
This corresponds to the beam step size used for our electron-beam lithography step that defines these features. 

Applying a variance to our model using only the bold parameters in Table~\ref{tab:tolerance analysis} results in an $f_0$ and $\Ql$ scatter given by the orange data of Figure~\ref{fig:scatter_data}. The model and measurement distributions are in excellent agreement, showing we can accurately model the fabrication errors.

\section{Full filter-bank simulations}\label{sec:filter-bank simulations}
\begin{figure}[htb]
    \centering
    \includegraphics[width=\textwidth]{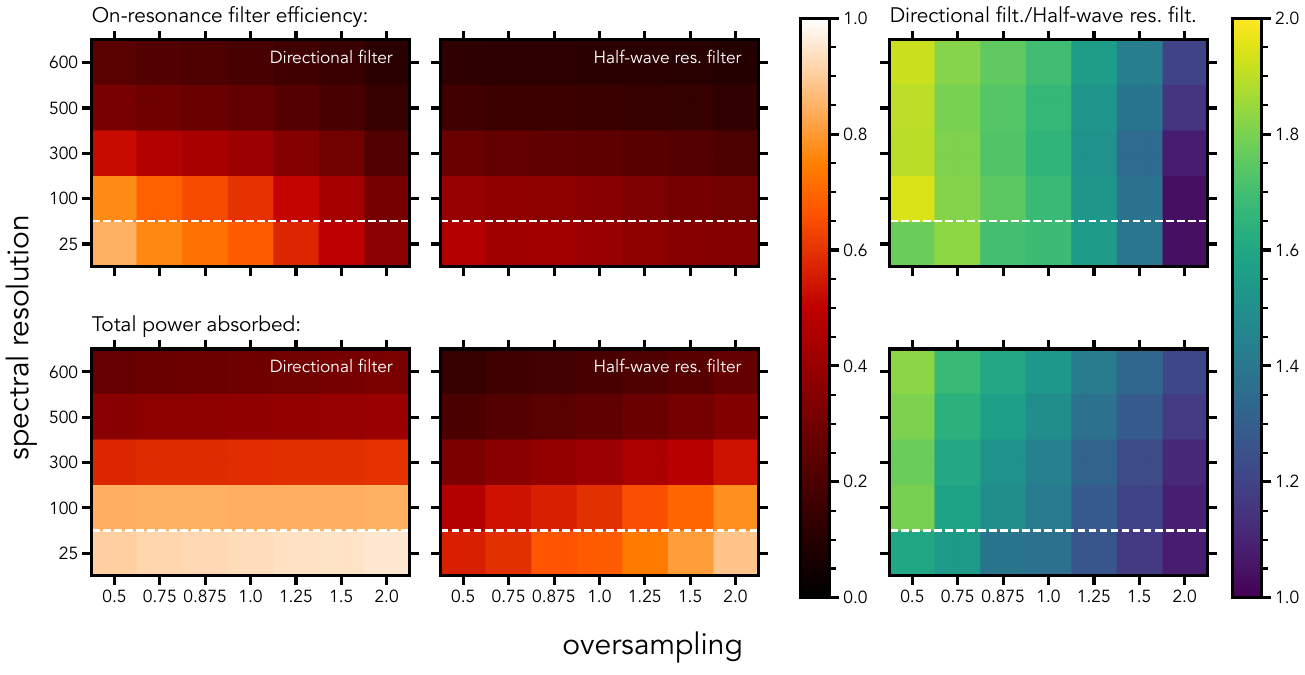}
    \caption{Simulations of realistic filter-banks (using the variations of the bold parameters in Table~\ref{tab:tolerance analysis}) across a range of spectral resolutions and oversampling values, using $\Qi = 1200$. The median values of both the on-resonance filter efficiency and the total power absorbed (also on-resonance) for each filter configuration are shown in the graph. The relative advantage of the directional filter is shown on the right.}
    \label{fig:optimization grid}
\end{figure}

We now simulate realistic filter-banks across a range of spectral resolutions and oversampling values and we compare the directional filter against the half-wave resonator filter. 
The realistic losses ($\Qi = 1200$) and fabrication errors (bold parameters in Table~\ref{tab:tolerance analysis}) are incorporated in this simulation. 
We note that in case of the directional filter the $\Qi$ and the $Z_0$ variance are applied independently to both resonators.
Figure~\ref{fig:optimization grid} shows the results.
The left-most column in each figure shows the near-isolated performance of both filter types for a range of spectral resolutions, since the undersampled filter-bank spaces the filters far apart in the spectrum. 
The bottom row is the case where fabrication tolerances and losses hardly matter due to the low spectral resolution, hence these can be seen as an ideal case with no scatter and no losses in the materials.

Figure~\ref{fig:optimization grid} shows that the directional filter outperforms the half-wave resonator filter, even with realistic losses and fabrication errors. 
From the rightmost panels in Figure~\ref{fig:optimization grid} it is clear that, for a given oversampling value, the performance ratio remains similar between the directional filter and the half-wave resonator filter going from a spectral resolution of 25, representing an ideal filter-bank without fabrication defects, to higher spectral resolutions, representing a realistic case.
In terms of on-resonance filter efficiency, the directional filter is close to being twice as efficient as the half-wave resonator filter for all spectral resolutions, showing that dielectric losses affect both filter types equally. Additionally we can see the effect of the limited $\Qi$ deteriorating the filter-bank performances in all cases for spectral resolutions above 100.

Only at high oversampling values does the median on-resonance filter efficiency of the directional filter converge to the efficiency of the half-wave resonator filter. 
At this point the filters are spectrally spaced so closely that the tails of neighboring filters are absorbing a significant amount of the power on-resonance.
However, even at the highest oversampling presented, the total power absorbed by the directional filter is still greater than that of the half-wave resonator filter. 
This means that even if one wants to sum the outputs of multiple filters to improve the total absorption, the directional filter configuration still remains the best choice.

Furthermore, the total power absorbed is near independent of oversampling for the directional filter. 
This means that the directional filter eliminates the need to oversample and sum outputs altogether \citep{hailey-dunsheathStatusSuperSpecBroadband2014}. 
The directional filter captures nearly all the power available after dielectric losses are accounted for, and it is the most efficient that any transmission line resonator filter can be.

\section{Conclusion}
We have introduced a phase-coherent filter, called a directional filter, as an alternative to the widely used shunted half-wave resonator filter for superconducting on-chip filter-banks. The directional filter overcomes the maximum pass-band response limit of \qty{50}{\%} for a half-wave resonator filter, with an efficiency of \qty{100}{\%} in a lossless scenario. This is achieved by matching the incoming and outgoing signals using a phase-coherent structure comprising two half-wave resonators.
In a filter-bank configuration the directional filter retains its advantage and has a nearly two times better efficiency compared to the half-wave resonator filter-bank.

Fabrication errors and dielectric losses are added to our model based on measurements from DESHIMA 2.0. We found that the frequency scatter is mainly caused by variations in the resonator characteristic impedance $Z_0$ due to nanometer scale line width variations.  We also found that the loaded quality factor $\Ql$ scatter can only be caused by internal quality factor $\Qi$ (dielectric loss) variations.

We can accurately represent a realistically fabricated filter-bank by incorporating these fabrication errors and dielectric losses for both filter types. 
Comparing these realistic filter-banks across a range of oversampling values and spectral resolutions shows that the directional filter performs better in all cases.

\backmatter

\section*{Acknowledgments}

The authors thank D.J. Thoen and L.G.G. Olde Scholtenhuis for their useful discussions on nanometer scale line width variations in e-beam lithography.

This work is supported by the European Union (ERC Consolidator Grant No. 101043486 TIFUUN). Views and opinions expressed are however those of the authors only and do not necessarily reflect those of the European Union or the European Research Council Executive Agency. Neither the European Union nor the granting authority can be held responsible for them.

\bibliography{bibliography}


\begin{thebibliography}{15}
\ifx \bisbn   \undefined \def \bisbn  #1{ISBN #1}\fi
\ifx \binits  \undefined \def \binits#1{#1}\fi
\ifx \bauthor  \undefined \def \bauthor#1{#1}\fi
\ifx \batitle  \undefined \def \batitle#1{#1}\fi
\ifx \bjtitle  \undefined \def \bjtitle#1{#1}\fi
\ifx \bvolume  \undefined \def \bvolume#1{\textbf{#1}}\fi
\ifx \byear  \undefined \def \byear#1{#1}\fi
\ifx \bissue  \undefined \def \bissue#1{#1}\fi
\ifx \bfpage  \undefined \def \bfpage#1{#1}\fi
\ifx \blpage  \undefined \def \blpage #1{#1}\fi
\ifx \burl  \undefined \def \burl#1{\textsf{#1}}\fi
\ifx \doiurl  \undefined \def \doiurl#1{\url{https://doi.org/#1}}\fi
\ifx \betal  \undefined \def \betal{\textit{et al.}}\fi
\ifx \binstitute  \undefined \def \binstitute#1{#1}\fi
\ifx \binstitutionaled  \undefined \def \binstitutionaled#1{#1}\fi
\ifx \bctitle  \undefined \def \bctitle#1{#1}\fi
\ifx \beditor  \undefined \def \beditor#1{#1}\fi
\ifx \bpublisher  \undefined \def \bpublisher#1{#1}\fi
\ifx \bbtitle  \undefined \def \bbtitle#1{#1}\fi
\ifx \bedition  \undefined \def \bedition#1{#1}\fi
\ifx \bseriesno  \undefined \def \bseriesno#1{#1}\fi
\ifx \blocation  \undefined \def \blocation#1{#1}\fi
\ifx \bsertitle  \undefined \def \bsertitle#1{#1}\fi
\ifx \bsnm \undefined \def \bsnm#1{#1}\fi
\ifx \bsuffix \undefined \def \bsuffix#1{#1}\fi
\ifx \bparticle \undefined \def \bparticle#1{#1}\fi
\ifx \barticle \undefined \def \barticle#1{#1}\fi
\bibcommenthead
\ifx \bconfdate \undefined \def \bconfdate #1{#1}\fi
\ifx \botherref \undefined \def \botherref #1{#1}\fi
\ifx \url \undefined \def \url#1{\textsf{#1}}\fi
\ifx \bchapter \undefined \def \bchapter#1{#1}\fi
\ifx \bbook \undefined \def \bbook#1{#1}\fi
\ifx \bcomment \undefined \def \bcomment#1{#1}\fi
\ifx \oauthor \undefined \def \oauthor#1{#1}\fi
\ifx \citeauthoryear \undefined \def \citeauthoryear#1{#1}\fi
\ifx \endbibitem  \undefined \def \endbibitem {}\fi
\ifx \bconflocation  \undefined \def \bconflocation#1{#1}\fi
\ifx \arxivurl  \undefined \def \arxivurl#1{\textsf{#1}}\fi
\csname PreBibitemsHook\endcsname

\bibitem[\protect\citeauthoryear{Endo et~al.}{2019}]{endoFirstLightDemonstration2019}
\begin{barticle}
\bauthor{\bsnm{Endo}, \binits{A.}},
\bauthor{\bsnm{Karatsu}, \binits{K.}},
\bauthor{\bsnm{Tamura}, \binits{Y.}},
\bauthor{\bsnm{Oshima}, \binits{T.}},
\bauthor{\bsnm{Taniguchi}, \binits{A.}},
\bauthor{\bsnm{Takekoshi}, \binits{T.}},
\bauthor{\bsnm{Asayama}, \binits{S.}},
\bauthor{\bsnm{Bakx}, \binits{T.J.L.C.}},
\bauthor{\bsnm{Bosma}, \binits{S.}},
\bauthor{\bsnm{Bueno}, \binits{J.}},
\bauthor{\bsnm{Chin}, \binits{K.W.}},
\bauthor{\bsnm{Fujii}, \binits{Y.}},
\bauthor{\bsnm{Fujita}, \binits{K.}},
\bauthor{\bsnm{Huiting}, \binits{R.}},
\bauthor{\bsnm{Ikarashi}, \binits{S.}},
\bauthor{\bsnm{Ishida}, \binits{T.}},
\bauthor{\bsnm{Ishii}, \binits{S.}},
\bauthor{\bsnm{Kawabe}, \binits{R.}},
\bauthor{\bsnm{Klapwijk}, \binits{T.M.}},
\bauthor{\bsnm{Kohno}, \binits{K.}},
\bauthor{\bsnm{Kouchi}, \binits{A.}},
\bauthor{\bsnm{Llombart}, \binits{N.}},
\bauthor{\bsnm{Maekawa}, \binits{J.}},
\bauthor{\bsnm{Murugesan}, \binits{V.}},
\bauthor{\bsnm{Nakatsubo}, \binits{S.}},
\bauthor{\bsnm{Naruse}, \binits{M.}},
\bauthor{\bsnm{Ohtawara}, \binits{K.}},
\bauthor{\bsnm{Pascual~Laguna}, \binits{A.}},
\bauthor{\bsnm{Suzuki}, \binits{J.}},
\bauthor{\bsnm{Suzuki}, \binits{K.}},
\bauthor{\bsnm{Thoen}, \binits{D.J.}},
\bauthor{\bsnm{Tsukagoshi}, \binits{T.}},
\bauthor{\bsnm{Ueda}, \binits{T.}},
\bauthor{\bsnm{{de Visser}}, \binits{P.J.}},
\bauthor{\bsnm{{van der Werf}}, \binits{P.P.}},
\bauthor{\bsnm{Yates}, \binits{S.J.C.}},
\bauthor{\bsnm{Yoshimura}, \binits{Y.}},
\bauthor{\bsnm{Yurduseven}, \binits{O.}},
\bauthor{\bsnm{Baselmans}, \binits{J.J.A.}}:
\batitle{First light demonstration of the integrated superconducting spectrometer}.
\bjtitle{Nat Astron}
\bvolume{3}(\bissue{11}),
\bfpage{989}--\blpage{996}
(\byear{2019})
\doiurl{10.1038/s41550-019-0850-8}
\end{barticle}
\endbibitem

\bibitem[\protect\citeauthoryear{Redford et~al.}{2022}]{redfordSuperSpecOnChipSpectrometer2022}
\begin{barticle}
\bauthor{\bsnm{Redford}, \binits{J.}},
\bauthor{\bsnm{Barry}, \binits{P.S.}},
\bauthor{\bsnm{Bradford}, \binits{C.M.}},
\bauthor{\bsnm{Chapman}, \binits{S.}},
\bauthor{\bsnm{Glenn}, \binits{J.}},
\bauthor{\bsnm{{Hailey-Dunsheath}}, \binits{S.}},
\bauthor{\bsnm{Janssen}, \binits{R.M.J.}},
\bauthor{\bsnm{Karkare}, \binits{K.S.}},
\bauthor{\bsnm{LeDuc}, \binits{H.G.}},
\bauthor{\bsnm{Mauskopf}, \binits{P.}},
\bauthor{\bsnm{McGeehan}, \binits{R.}},
\bauthor{\bsnm{Shirokoff}, \binits{E.}},
\bauthor{\bsnm{Wheeler}, \binits{J.}},
\bauthor{\bsnm{Zmuidzinas}, \binits{J.}}:
\batitle{{{SuperSpec}}: {{On-Chip Spectrometer Design}}, {{Characterization}}, and {{Performance}}}.
\bjtitle{J Low Temp Phys}
\bvolume{209}(\bissue{3}),
\bfpage{548}--\blpage{555}
(\byear{2022})
\doiurl{10.1007/s10909-022-02866-x}
\end{barticle}
\endbibitem

\bibitem[\protect\citeauthoryear{Cataldo et~al.}{2019}]{cataldoSecondgenerationMicroSpecCompact2019}
\begin{barticle}
\bauthor{\bsnm{Cataldo}, \binits{G.}},
\bauthor{\bsnm{Barrentine}, \binits{E.M.}},
\bauthor{\bsnm{Bulcha}, \binits{B.T.}},
\bauthor{\bsnm{Ehsan}, \binits{N.}},
\bauthor{\bsnm{Hess}, \binits{L.A.}},
\bauthor{\bsnm{Noroozian}, \binits{O.}},
\bauthor{\bsnm{Stevenson}, \binits{T.R.}},
\bauthor{\bsnm{Wollack}, \binits{E.J.}},
\bauthor{\bsnm{Moseley}, \binits{S.H.}},
\bauthor{\bsnm{Switzer}, \binits{E.R.}}:
\batitle{Second-generation {{Micro-Spec}}: {{A}} compact spectrometer for far-infrared and submillimeter space missions}.
\bjtitle{Acta Astronautica}
\bvolume{162},
\bfpage{155}--\blpage{159}
(\byear{2019})
\doiurl{10.1016/j.actaastro.2019.06.012}
\end{barticle}
\endbibitem

\bibitem[\protect\citeauthoryear{Pascual~Laguna et~al.}{2021}]{pascuallagunaTerahertzBandPassFilters2021}
\begin{barticle}
\bauthor{\bsnm{Pascual~Laguna}, \binits{A.}},
\bauthor{\bsnm{Karatsu}, \binits{K.}},
\bauthor{\bsnm{Thoen}, \binits{D.J.}},
\bauthor{\bsnm{Murugesan}, \binits{V.}},
\bauthor{\bsnm{Buijtendorp}, \binits{B.T.}},
\bauthor{\bsnm{Endo}, \binits{A.}},
\bauthor{\bsnm{Baselmans}, \binits{J.J.A.}}:
\batitle{Terahertz {{Band-Pass Filters}} for {{Wideband Superconducting On-Chip Filter-Bank Spectrometers}}}.
\bjtitle{IEEE Transactions on Terahertz Science and Technology}
\bvolume{11}(\bissue{6}),
\bfpage{635}--\blpage{646}
(\byear{2021})
\doiurl{10.1109/TTHZ.2021.3095429}
\end{barticle}
\endbibitem

\bibitem[\protect\citeauthoryear{Kov{\'a}cs et~al.}{2012}]{kovacsSuperSpecDesignConcept2012}
\begin{bchapter}
\bauthor{\bsnm{Kov{\'a}cs}, \binits{A.}},
\bauthor{\bsnm{Barry}, \binits{P.S.}},
\bauthor{\bsnm{Bradford}, \binits{C.M.}},
\bauthor{\bsnm{Chattopadhyay}, \binits{G.}},
\bauthor{\bsnm{Day}, \binits{P.}},
\bauthor{\bsnm{Doyle}, \binits{S.}},
\bauthor{\bsnm{{Hailey-Dunsheath}}, \binits{S.}},
\bauthor{\bsnm{Hollister}, \binits{M.}},
\bauthor{\bsnm{McKenney}, \binits{C.}},
\bauthor{\bsnm{LeDuc}, \binits{H.G.}},
\bauthor{\bsnm{Llombart}, \binits{N.}},
\bauthor{\bsnm{Marrone}, \binits{D.P.}},
\bauthor{\bsnm{Mauskopf}, \binits{P.}},
\bauthor{\bsnm{O'Brient}, \binits{R.C.}},
\bauthor{\bsnm{Padin}, \binits{S.}},
\bauthor{\bsnm{Swenson}, \binits{L.J.}},
\bauthor{\bsnm{Zmuidzinas}, \binits{J.}}:
\bctitle{{{SuperSpec}}: Design concept and circuit simulations}.
In: \bbtitle{Millimeter, {{Submillimeter}}, and {{Far-Infrared~Detectors}} and {{Instrumentation}} For {{Astronomy VI}}},
vol. \bseriesno{8452},
pp. \bfpage{748}--\blpage{757}
(\byear{2012}).
\doiurl{10.1117/12.927160}
\end{bchapter}
\endbibitem

\bibitem[\protect\citeauthoryear{Endo et~al.}{2012}]{endoDesignIntegratedFilterbank2012}
\begin{barticle}
\bauthor{\bsnm{Endo}, \binits{A.}},
\bauthor{\bsnm{{van~der Werf}}, \binits{P.}},
\bauthor{\bsnm{Janssen}, \binits{R.M.J.}},
\bauthor{\bsnm{{de Visser}}, \binits{P.J.}},
\bauthor{\bsnm{Klapwijk}, \binits{T.M.}},
\bauthor{\bsnm{Baselmans}, \binits{J.J.A.}},
\bauthor{\bsnm{Ferrari}, \binits{L.}},
\bauthor{\bsnm{Baryshev}, \binits{A.M.}},
\bauthor{\bsnm{Yates}, \binits{S.J.C.}}:
\batitle{Design of an {{Integrated Filterbank}} for {{DESHIMA}}: {{On-Chip Submillimeter Imaging Spectrograph Based}} on {{Superconducting Resonators}}}.
\bjtitle{J Low Temp Phys}
\bvolume{167}(\bissue{3}),
\bfpage{341}--\blpage{346}
(\byear{2012})
\doiurl{10.1007/s10909-012-0502-1}
\end{barticle}
\endbibitem

\bibitem[\protect\citeauthoryear{Taniguchi et~al.}{2022}]{taniguchiDESHIMADevelopmentIntegrated2022}
\begin{barticle}
\bauthor{\bsnm{Taniguchi}, \binits{A.}},
\bauthor{\bsnm{Bakx}, \binits{T.J.L.C.}},
\bauthor{\bsnm{Baselmans}, \binits{J.J.A.}},
\bauthor{\bsnm{Huiting}, \binits{R.}},
\bauthor{\bsnm{Karatsu}, \binits{K.}},
\bauthor{\bsnm{Llombart}, \binits{N.}},
\bauthor{\bsnm{Rybak}, \binits{M.}},
\bauthor{\bsnm{Takekoshi}, \binits{T.}},
\bauthor{\bsnm{Tamura}, \binits{Y.}},
\bauthor{\bsnm{Akamatsu}, \binits{H.}},
\bauthor{\bsnm{Brackenhoff}, \binits{S.}},
\bauthor{\bsnm{Bueno}, \binits{J.}},
\bauthor{\bsnm{Buijtendorp}, \binits{B.T.}},
\bauthor{\bsnm{Dabironezare}, \binits{S.O.}},
\bauthor{\bsnm{Doing}, \binits{A.-K.}},
\bauthor{\bsnm{Fujii}, \binits{Y.}},
\bauthor{\bsnm{Fujita}, \binits{K.}},
\bauthor{\bsnm{Gouwerok}, \binits{M.}},
\bauthor{\bsnm{H{\"a}hnle}, \binits{S.}},
\bauthor{\bsnm{Ishida}, \binits{T.}},
\bauthor{\bsnm{Ishii}, \binits{S.}},
\bauthor{\bsnm{Kawabe}, \binits{R.}},
\bauthor{\bsnm{Kitayama}, \binits{T.}},
\bauthor{\bsnm{Kohno}, \binits{K.}},
\bauthor{\bsnm{Kouchi}, \binits{A.}},
\bauthor{\bsnm{Maekawa}, \binits{J.}},
\bauthor{\bsnm{Matsuda}, \binits{K.}},
\bauthor{\bsnm{Murugesan}, \binits{V.}},
\bauthor{\bsnm{Nakatsubo}, \binits{S.}},
\bauthor{\bsnm{Oshima}, \binits{T.}},
\bauthor{\bsnm{Pascual~Laguna}, \binits{A.}},
\bauthor{\bsnm{Thoen}, \binits{D.J.}},
\bauthor{\bsnm{{van der Werf}}, \binits{P.P.}},
\bauthor{\bsnm{Yates}, \binits{S.J.C.}},
\bauthor{\bsnm{Endo}, \binits{A.}}:
\batitle{{{DESHIMA}} 2.0: {{Development}} of an {{Integrated Superconducting Spectrometer}} for {{Science-Grade Astronomical Observations}}}.
\bjtitle{J Low Temp Phys}
\bvolume{209}(\bissue{3}),
\bfpage{278}--\blpage{286}
(\byear{2022})
\doiurl{10.1007/s10909-022-02888-5}
\end{barticle}
\endbibitem

\bibitem[\protect\citeauthoryear{Cohn and Coale}{1956}]{cohnDirectionalChannelSeparationFilters1956}
\begin{barticle}
\bauthor{\bsnm{Cohn}, \binits{S.B.}},
\bauthor{\bsnm{Coale}, \binits{F.S.}}:
\batitle{Directional {{Channel-Separation Filters}}}.
\bjtitle{Proceedings of the IRE}
\bvolume{44}(\bissue{8}),
\bfpage{1018}--\blpage{1024}
(\byear{1956})
\doiurl{10.1109/JRPROC.1956.275043}
\end{barticle}
\endbibitem

\bibitem[\protect\citeauthoryear{Coale}{1956}]{coaleTravelingWaveDirectionalFilter1956}
\begin{barticle}
\bauthor{\bsnm{Coale}, \binits{F.S.}}:
\batitle{A {{Traveling-Wave Directional Filter}}}.
\bjtitle{IRE Transactions on Microwave Theory and Techniques}
\bvolume{4}(\bissue{4}),
\bfpage{256}--\blpage{260}
(\byear{1956})
\doiurl{10.1109/TMTT.1956.1125073}
\end{barticle}
\endbibitem

\bibitem[\protect\citeauthoryear{Oliver}{1954}]{oliverDirectionalElectromagneticCouplers1954}
\begin{barticle}
\bauthor{\bsnm{Oliver}, \binits{B.M.}}:
\batitle{Directional {{Electromagnetic Couplers}}}.
\bjtitle{Proceedings of the IRE}
\bvolume{42}(\bissue{11}),
\bfpage{1686}--\blpage{1692}
(\byear{1954})
\doiurl{10.1109/JRPROC.1954.274664}
\end{barticle}
\endbibitem

\bibitem[\protect\citeauthoryear{Jones}{1956}]{jonesCoupledStripTransmissionLineFiltersDirectional1956}
\begin{barticle}
\bauthor{\bsnm{Jones}, \binits{E.M.T.}}:
\batitle{Coupled-{{Strip-Transmission-Line Filters}} and {{Directional Couplers}}}.
\bjtitle{IRE Transactions on Microwave Theory and Techniques}
\bvolume{4}(\bissue{2}),
\bfpage{75}--\blpage{81}
(\byear{1956})
\doiurl{10.1109/TMTT.1956.1125022}
\end{barticle}
\endbibitem

\bibitem[\protect\citeauthoryear{Tuttle and Wanselow}{1959}]{tuttlePracticalDesignStripTransmissionLine1959}
\begin{barticle}
\bauthor{\bsnm{Tuttle}, \binits{L.P.}},
\bauthor{\bsnm{Wanselow}, \binits{R.D.}}:
\batitle{Practical {{Design}} of {{Strip-Transmission-Line Half-Wavelength Resonator Directional Filters}}}.
\bjtitle{IRE Transactions on Microwave Theory and Techniques}
\bvolume{7}(\bissue{1}),
\bfpage{168}--\blpage{173}
(\byear{1959})
\doiurl{10.1109/TMTT.1959.1124642}
\end{barticle}
\endbibitem

\bibitem[\protect\citeauthoryear{Janssen et~al.}{2013}]{janssenHighOpticalEfficiency2013}
\begin{barticle}
\bauthor{\bsnm{Janssen}, \binits{R.M.J.}},
\bauthor{\bsnm{Baselmans}, \binits{J.J.A.}},
\bauthor{\bsnm{Endo}, \binits{A.}},
\bauthor{\bsnm{Ferrari}, \binits{L.}},
\bauthor{\bsnm{Yates}, \binits{S.J.C.}},
\bauthor{\bsnm{Baryshev}, \binits{A.M.}},
\bauthor{\bsnm{Klapwijk}, \binits{T.M.}}:
\batitle{High optical efficiency and photon noise limited sensitivity of microwave kinetic inductance detectors using phase readout}.
\bjtitle{Appl. Phys. Lett.}
\bvolume{103}(\bissue{20}),
\bfpage{203503}
(\byear{2013})
\doiurl{10.1063/1.4829657}
\end{barticle}
\endbibitem

\bibitem[\protect\citeauthoryear{{Hailey-Dunsheath} et~al.}{2014}]{hailey-dunsheathStatusSuperSpecBroadband2014}
\begin{bchapter}
\bauthor{\bsnm{{Hailey-Dunsheath}}, \binits{S.}},
\bauthor{\bsnm{Shirokoff}, \binits{E.}},
\bauthor{\bsnm{Barry}, \binits{P.S.}},
\bauthor{\bsnm{Bradford}, \binits{C.M.}},
\bauthor{\bsnm{Chattopadhyay}, \binits{G.}},
\bauthor{\bsnm{Day}, \binits{P.}},
\bauthor{\bsnm{Doyle}, \binits{S.}},
\bauthor{\bsnm{Hollister}, \binits{M.}},
\bauthor{\bsnm{Kovacs}, \binits{A.}},
\bauthor{\bsnm{LeDuc}, \binits{H.G.}},
\bauthor{\bsnm{Mauskopf}, \binits{P.}},
\bauthor{\bsnm{McKenney}, \binits{C.M.}},
\bauthor{\bsnm{Monroe}, \binits{R.}},
\bauthor{\bsnm{O'Brient}, \binits{R.}},
\bauthor{\bsnm{Padin}, \binits{S.}},
\bauthor{\bsnm{Reck}, \binits{T.}},
\bauthor{\bsnm{Swenson}, \binits{L.}},
\bauthor{\bsnm{Tucker}, \binits{C.E.}},
\bauthor{\bsnm{Zmuidzinas}, \binits{J.}}:
\bctitle{Status of {{SuperSpec}}: A broadband, on-chip millimeter-wave spectrometer}.
In: \bbtitle{Millimeter, {{Submillimeter}}, and {{Far-Infrared~Detectors}} and {{Instrumentation}} for {{Astronomy VII}}},
vol. \bseriesno{9153},
pp. \bfpage{181}--\blpage{196}
(\byear{2014}).
\doiurl{10.1117/12.2057229}
\end{bchapter}
\endbibitem

\bibitem[\protect\citeauthoryear{Pascual~Laguna}{2022}]{pascuallagunaOnChipSolutionsFuture2022}
\begin{botherref}
\oauthor{\bsnm{Pascual~Laguna}, \binits{A.}}:
On-{{Chip Solutions}} for {{Future THz Imaging Spectrometers}}.
PhD thesis,
Delft University of Technology
(2022).
\doiurl{10.4233/uuid:512da833-dc9d-4301-b566-35243d2f1f9b}
\end{botherref}
\endbibitem

\end{thebibliography}

\end{document}